\def\Title{Classical Three-Box ``paradox''}
\def\arXiv{arXiv: quant-ph/0207124 v4}

\def\Abstract{%
A simple classical probabilistic system (a simple card game) classically exemplifies Aharonov and Vaidman's ``Three-Box `paradox'\,'' [J.~Phys.~A {\bf24}, 2315 (1991)], implying that the Three-Box example is neither quantal nor a paradox and leaving one less difficulty to busy the interpreters of quantum mechanics. An ambiguity in the usual expression of the retrodiction formula is shown to have misled Albert, Aharonov, and D'Amato [Phys.~Rev.~Lett.~{\bf54}, 5 (1985)] to a result not, in fact, ``curious''; the discussion illustrates how to avoid this ambiguity.
}%
\documentclass[aps,rmp,nobibnotes,nofootinbib,onecolumn,twoside,noshowpacs,draft]{revtex4}%
 \usepackage[centertags,sumlimits]{amsmath}
 \usepackage{amssymb}
 \makeatletter
 \def\p@section{}
 \def\p@subsection{}
 \def\p@subsubsection{}
 \def\p@paragraph{}
 \def\p@subparagraph{}

 \def\@hangfrom@section#1#2#3{\@hangfrom{#1}{\large\textrm{#2}}{\large\textrm{#3}}}%
 \def\@hangfrom@subsection#1#2#3{\@hangfrom{#1}{\textrm{#2}}{\textrm{#3}}}%
 \def\@hangfrom@subsubsection#1#2#3{\@hangfrom{#1}{\textrm{#2}}{\textrm{#3}}}%
 \def\frontmatter@setup{\normalfont\raggedright}% Not san serif now
 \makeatother

 \usepackage{xspace}

 \newcommand{\ie}{i.e., }
 \newcommand{\eg}{e.g., }

 \newcommand{\cf}{cf.\xspace}

 \newcommand{\RefSec}[1]{Sec.~\ref{#1}}

 \newcommand{\RefEqn}[1]{\eqref{#1}}
 \newcommand{\RefEqns}[1]{Eqs.~\eqref{#1}}

 \newcommand{\abs}[1]{\ensuremath{\left\vert#1\right\vert}}
 \newcommand{\Sum}[1]{\ensuremath{\sum_{#1}}}
 \newcommand{\KDelta}[2]{\ensuremath{\delta_{{#1}{#2}}}}
 \newcommand{\set}[1]{\ensuremath{{\left\{\,#1\,\right\}}}}%

 \newcommand{\Or}{\ensuremath{\,\vee\,}}
 \newcommand{\Orj}[2]{\ensuremath{{\textstyle\bigvee}_{\!#1}\,{#2}}}

 \newcommand{\One}{\ensuremath{\mathbf{\displaystyle{1}}}\xspace}

 \newcommand{\Sys}[1]{\ensuremath{\mathcal{#1}}\xspace}

 \newcommand{\Trace}[2][]{\ensuremath{{\rm Tr}%
                    {\!}_{\Sys{#1}}\Bigl\{\,{#2}\Bigr\}}}

 \newcommand{\ket}[1]{\ensuremath{\vert\,{#1}\,\rangle}}
 \newcommand{\bra}[1]{\ensuremath{\langle\,#1\,\vert}}
 \newcommand{\braket}[2]{\ensuremath{\langle#1\,\vert\,#2\rangle}}% <#1|#2>

 \newcommand{\ProjSym}{\boldsymbol{\mathsf{P}}}
 \newcommand{\Proj}[2][]{\ensuremath{\ProjSym^{\Sys{#1}}[\,#2\,]}\xspace}

 \newcommand{\bRho}{\pmb{\rho}}

 \renewcommand{\Pr}[2][]{\ensuremath{{\rm Pr}_{#1}\bigl(\,{#2}\,\bigr)}}
 \newcommand{\Prob}[3][]{\ensuremath{{\rm Pr}_{#1}\bigl(\,{#2}\bigm|#3\,\bigr)}}

 \newcommand{\p}[1]{\ensuremath{p_{#1}}\xspace}
 \newcommand{\q}[1]{\ensuremath{q_{#1}}\xspace}

 \newcommand{\Mem}{\ensuremath{\mathcal{M}}\xspace}
 \newcommand{\Face}{\ensuremath{\text{\textsl{Face}}}\xspace}
 \newcommand{\Suit}{\ensuremath{\text{\textsl{Suit}}}\xspace}
 \newcommand{\K}{\ensuremath{\text{\textsf{K}}}\xspace}
 \newcommand{\Q}{\ensuremath{\text{\textsf{Q}}}\xspace}
 \newcommand{\J}{\ensuremath{\text{\textsf{J}}}\xspace}
 
 \renewcommand{\S}{\ensuremath{\text{\textsf{S}}}\xspace}
 \renewcommand{\H}{\ensuremath{\text{\textsf{H}}}\xspace}
 \newcommand{\D}{\ensuremath{\text{\textsf{D}}}\xspace}
 \newcommand{\These}{\ensuremath{\text{\textsl{These}}}\xspace}
 \newcommand{\Others}{\ensuremath{\text{\textsl{Others}}}\xspace}

 \newcommand{\MP}{\ensuremath{M_P}\xspace}
 \newcommand{\MPs}[1]{\ensuremath{M_P^{[#1]}}\xspace}

 \newcommand{\ps}[2]{\ensuremath{p_{#2}^{[#1]}}\xspace}
 \newcommand{\qs}[2]{\ensuremath{q_{#2}^{[#1]}}\xspace}

 \newcommand{\sz}{s^{[0]}}

 \newcommand{\Mp}[2]{\ensuremath{M_{#1_{#2}}^{\sim}}\xspace}
 \newcommand{\MPj}{\ensuremath{\Mp{P}{j}}\xspace}
 \newcommand{\MPjOne}{{\MPj}^{[1]}}

 \newcommand{\qq}{\ensuremath{q^{[2]}}\xspace}

 \newcommand{\psneg}[2]{\ensuremath{{p_{#2}^{\sim}}^{[#1]}}\xspace}

\begin{document}
 \makeatletter
 \def\ps@titlepage{%
   \renewcommand{\@oddfoot}{\parbox{\textwidth}{%
     \vglue-5em Journal of Physics A: Mathematical and General 
                \copyright 2003 IOP Publishing Ltd.\\
     Online at stacks.iop.org/JPhysA/36/4891}}%
   \renewcommand{\@evenfoot}{}%
   \renewcommand{\@oddhead}{\emph{J.~Phys.~A} {\bf36}(17) 4891-4900 (2003)%
                             \quad(corrected)\hfill\arXiv}
   \renewcommand{\@evenhead}{}}
 \makeatother

%%%%%%%%%%%%%%%%%%%%%%%%%%%%%%%%%%%%%%%%%%%%%%%%%%%%%%%%%%%%%%%%%%%%%%%%%%%%%%%%%%%%%%%%
\title[Kirkpatrick -- \Title] %% for running titles on pages
      {\Title} %% the title-page title
%%%%%% PERSONAL %%%%%%%%%%%%%%%%%%%%%%%%%%%%%%%%%%%%%%%%%%%%%%%%%%%%%%%%%%%%%%%%%%%%%%%%
\author{K.~A.~Kirkpatrick}
\email[E-mail: ]{kirkpatrick@physics.nmhu.edu}
\affiliation{New Mexico Highlands
University, Las Vegas, New Mexico 87701}
%%%%%%%%%%%%%%%%%%%%%%%%%%%%%%%%%%%%%%%%%%%%%%%%%%%%%%%%%%%%%%%%%%%%%%%%%%%%%%%%%%%%%%%%
%%%%%% ABSTRACT %%%%%%%%%%%%%%%%%%%%%%%%%%%%%%%%%%%%%%%%%%%%%%%%%%%%%%%%%%%%%%%%%%%%%%%%
\begin{abstract}
\Abstract
\end{abstract}
%%%%%%%%%%%%%%%%%%%%%%%%%%%%%%%%%%%%%%%%%%%%%%%%%%%%%%%%%%%%%%%%%%%%%%%%%%%%%%%%%%%%%%%%
 \maketitle
%%%%%%%%%%%%%%%%%%%%%%%%%%%%%%%%%%%%%%%%%%%%%%%%%%%%%%%%%%%%%%%%%%%%%%%%%%%%%%%%%%%%%%%%
%% centered short title in each header:
 \makeatletter\markboth{\hfill\@shorttitle\hfill}{\hfill\@shorttitle\hfill}\makeatother
 \pagestyle{myheadings}
%%%%%%%%%%%%%%%%%%%%%%%%%%%%%%%%%%%%%%%%%%%%%%%%%%%%%%%%%%%%%%%%%%%%%%%%%%%%%%%%%%%%%%%%
%%% BODY OF DOCUMENT %%%%%%%%%%%%%%%%%%%%%%%%%%%%%%%%%%%%%%%%%%%%%%%%%%%%%%%%%%%%%%%%%%%
%%%%%%%%%%%%%%%%%%%%%%%%%%%%%%%%%%%%%%%%%%%%%%%%%%%%%%%%%%%%%%%%%%%%%%%%%%%%%%%%%%%%%%%%
\section{Introduction}%
\nocite{AharonovV91} Aharonov and Vaidman (1991; henceforth ``AV'') introduced what has come to be called the ``Three-Box paradox,'' a postselected process in which each of two disjoint events occurs with certainty. They express this example in terms of a particle and three boxes; the process starts with the particle in a state smeared over all three boxes and is postselected to end with the particle in another, similarly smeared, state. These terminal states are chosen so that, if the first of the boxes is opened during the process, the particle is certain to be found there, while if the second box is opened, the particle is found there: a single particle is certain to be found in each of two boxes! \textcite{Vaidman96} sums this up with
\begin{quote}
The elements of reality for pre- and post\-selected quantum systems have unusual and counterintuitive properties. But, may be this is not because of the illness of their definition, but due to bizarre feature of quantum systems which goes against the intuition built during thousands of years, when the results of quantum experiments were not known.
\end{quote}
But this, I think, is much ado about nothing. I will present an example of this behavior in a classical-probability setting. (The example is in terms of playing cards, rather than particles in boxes; the paradoxical result will be that the card drawn must at one and the same time be a Diamond and a Spade.) Because the setting of this system is so ordinary, we cannot be tempted to dismiss its behavior as another ``bizarre feature of quantum mechanics''---instead, we are led to look for the misapprehension which has led us astray. We certainly will not accede to the proposed identification of ``elements of reality'' with ``certainty of occurrence''---the elements of reality of this classical system are quite visible and identifiable, not easily conflated with ghosts.

In \RefSec{S:ThreeBox}, I present AV's Three-Box example, restate it in terms of the physically less obscure triple-slit atomic Young device, and analyze it in classical probability terms. In \RefSec{S:theSystem} I present a classical system (a deck of cards) with probabilistic properties which allow the construction of a classical Three-Box system.

Finally, in \RefSec{S:Curious}, I describe how an ambiguity in the presentation by Aharonov, Bergmann, and Lebowitz (1964; henceforth ``ABL'')\nocite{ABL} of the retrodiction formula misled Albert, Aharonov, and D'Amato (1985) \nocite{AlbertAD85} to a result they thought ``Curious'' (which, in turn, misled AV to the Three-Box example).

\section{The Three-Box system and its ``paradox''}\label{S:ThreeBox}
The ``Three-Box paradox'' of AV consists of a single particle and three boxes; the value $\p{j}$ of the observable $P$ denote the presence of the particle in the corresponding box $j$ ($j=1..3$).  The system is prepared in a certain state $s$ at the time $t_0$, and only those occurrences for which the system is detected, at time $t_1$, to be in a particular state $q$ are considered. At an intermediate time $t\in(t_0,\,t_1)$ we may look for the particle by opening a box. AV showed that with
\begin{equation}\label{E:threeboxQconditionEG}
 \ket{s}=(\ket{\p{1}}+\ket{\p{2}}+\ket{\p{3}})/\sqrt{3}\quad\text{and}\quad%
 \ket{q}=(\ket{\p{1}}+\ket{\p{2}}-\ket{\p{3}})/\sqrt{3},
\end{equation}
if we open box~1, we (always) find the particle in it; however, they showed, it is also true that if we instead open box~2, we (always) find the particle in \emph{it}: The retrodictive observation of \p{1} is certain, and the retrodictive observation of \p{2} is certain---the Three-Box ``paradox.''

\subsection{Three-slit Young implementation of the Three-Box system}\label{SS:ThreeSlit}%
The reader may find the physical significance of the initial and final states of the Three-Box system rather obscure, expressed in terms of particles in boxes. These states, as well as the entire system, are much more easily understood expressed as a three-slit atomic diffraction system. Expressing the Three-Box system in this form emphasizes that its paradoxical behavior is as ordinary (to the extent that any atomic Young system may be considered ``ordinary''!) as a double-slit interference apparatus.

Three slits are equally spaced with a separation $a$. The top and bottom slits are labeled 1 and 2, and the middle slit, 3. A detector $D$ is placed on-axis, at a distance $L$ from slit~3 so that $\sqrt{L^2+a^2}-L=\lambda/2$ (with $\lambda$ the wavelength). A detector $d$ is placed at slit~1 or slit~2 (for atoms, a micromaser; for photons, a quarter-wave plate, the photon source linearly polarized). This system's initial and final states are described by \RefEqn{E:threeboxQconditionEG}, and it behaves exactly as the Three-Box example: If $d$ is placed at slit~1, every detection at $D$ is in coincidence with a detection at $d$ (implying passage through slit~1); if $d$ is placed at slit~2, every detection at $D$ is in coincidence with a detection at $d$ (implying passage through slit~2). This behavior is easily understood: Placing a detector at slit~1 creates the disjunction ``either the particle passed through slit~1 or it passed through the double-slit apparatus comprised of slits~2 and~3''; but passage through the double-slit destructively interferes at the detector $D$, so the second term of the disjunction must be false, forcing the first to be true. This apparatus is symmetric under exchange of slits~1 and 2---hence the ``paradox.''

The three slits correspond to the three boxes; a detector at slit~1 (only) corresponds to the opening of box~1 (only). The ``paradox'' of the Three-Box system reduces to the phenomenon (paradox?) of destructive interference in a two-slit Young apparatus, nothing more.

\subsection{Classical probability derivation of the Three-Box retrodiction
formula}\label{S:ClassDeriv}%
Let us now derive the expression for the retrodictive probability applicable to the Three-Box example. We carry out the derivation in classical probability terms to avoid being misled by any quantum ``bizarreness.'' (See Appendix \ref{S:Notation} for notational matters.)

The observation of the contents of a single box is a \emph{partial} observation: The opening of box~1, for example, determines ``$\p{1}\vee\p{1}^{\sim}$,'' where \p{1} denotes ``particle in box~1'' and $\p{1}^{\sim}$ denotes ``particle not in box~1.'' Note that $\p{1}^{\sim}$ is not the same as $\p{2}\vee\p{3}$; $\p{1}^{\sim}$ does not signify the ignorance of ``either \p{2} or \p{3}, but we don't know which,'' but rather signifies the lack of a fact of the matter regarding these two possibilities.%
\footnote{%
Aristotle gave this example: If $P$ is the proposition ``Tomorrow there will be a sea battle,'' then $P\vee{P}^{\sim}$ is true, but neither $P$ nor ${P}^{\sim}$ has a truth value today---there is no fact of the matter regarding either.
} %
The condition of such a partial observation is the \emph{partial manifestation}
\begin{equation}
  \MPj\equiv\p{j}\vee\p{j}^{\sim};
\end{equation}
of ``in box~$j$'' and ``not in box~$j$,'' exactly \emph{one} is true.

Accounting for this manifestation, and carefully labeling each event's position in the sequence with a bracketed superscript ordinal (see Appendix \ref{S:Notation}), the expression for the retrodictive probability of finding the particle when box $j$ is opened is
\begin{equation}\label{E:FirstStep}
 \Prob[\sz]{\ps{1}{j}}{\MPjOne\wedge\qs{2}{k}}%
  =\frac%
    {\Prob[\sz]{\ps{1}{j}\wedge\qs{2}{k}}{\MPjOne}}%
    {\Prob[\sz]{\qs{2}{k}}{\MPjOne}},
\end{equation}
where we have used \RefEqn{E:Cond}. (Note that, were the manifestation not expressed explicitly, the denominator would be ambiguous as to to the identity or nature of $\text{event}^{[1]}$.)

Because the condition $\p{j}\wedge\MPj=\p{j}$, the numerator is
\begin{equation}\label{E:noManif}
  \Prob[\sz]{\ps{1}{j}\wedge\qs{2}{k}}{\MPjOne}=
  \Prob[\sz]{\qs{2}{k}}{\ps{1}{j}}\,\Pr[\sz]{\ps{1}{j}}.
\end{equation}

Further, the condition $\MPj\wedge(\p{j}\vee\p{j}^{\sim})=\MPj$; thus the denominator may be written
\begin{align}
 &\Prob[\sz]{\qs{2}{k}}{\MPjOne}=%
    \frac{\Prob[\sz]{\big(\p{j}\vee\p{j}^{\sim}\big)^{[1]}\wedge\qs{2}{k}}{\MPjOne}}%
         {\Pr[\sz]{\big(\p{j}\vee\p{j}^{\sim}\big)^{[1]}}}\notag\\
 &\qquad=\Prob[\sz]{\ps{1}{j}\wedge\qs{2}{k}}{\MPjOne}+%\notag\\
   \Prob[\sz]{\psneg{1}{j}\wedge\qs{2}{k}}{\MPjOne}\notag\\
 &\qquad=\Prob[s]{\q{k}}{\p{j}}\,\Pr[s]{\p{j}}+%
   \Prob[s]{\q{k}}{\p{j}^{\sim}}\,\Pr[s]{\p{j}^{\sim}},
\end{align}
where we have used \RefEqn{E:noManif} as well as the disjointness and completeness of \set{\p{j},\,{\p{j}}^{\sim}}. Thus the retrodictive probability \RefEqn{E:FirstStep} may be written
\begin{equation}\label{E:threebox}
  \Prob[\sz]{\ps{1}{j}}{\MPjOne\wedge\qq}=\frac%
    {\Prob[s]{q}{\p{j}}\Pr[s]{\p{j}}}%
    {\Prob[s]{q}{\p{j}}\Pr[s]{\p{j}}+\Prob[s]{q}{{\p{j}^{\sim}}}\Pr[s]{\p{j}^{\sim}}}.
\end{equation}

\subsection{The Three-Box paradox}%
If the second term of the denominator of \RefEqn{E:threebox} were to vanish, the probability would be 1; if there is a choice of $s$ and $q$ such that term were to vanish for more than one value of $j$ (\eg for both \Mp{P}{1} and \Mp{P}{2}), we would obtain a Three-Box ``paradox.'' In the succeeding section we will present a classical system which exhibits exactly this behavior.

The Three-Box result requires \emph{partial} manifestation; if, instead, we make a complete observation $\MP=\p{1}\vee\p{2}\vee\p{3}$ (\eg by looking into at least two boxes), then the manifestation is complete,
\begin{equation}\label{E:RetroA}
\Prob[\sz]{\ps{1}{j}}{\MPs{1}\wedge\qs{2}{k}}=\\
  \frac{\Prob[s]{\q{k}}{\p{j}}\Pr[s]{\p{j}}}%
         {\sum_{t=1}^3\Prob[s]{\q{k}}{\p{t}}\Pr[s]{\p{t}}}.
\end{equation}
From this it is clear that $\Prob[\sz]{\ps{1}{j}}{\MP^{[1]}\wedge\qq}=1$ is not possible for more than one $j$, no matter the choice of $s$ and $q$.

\subsection{The quantal Three-Box paradox}%
If the particle is first placed in box~1 and then box~2 is opened (of course it is not there), it should be that if box~3 were next opened, the particle would not be there, but if, instead, box~1 were next opened, the particle \emph{would} be found. This is the requirement of \emph{stability}, \RefEqn{E:Stability}: the act of looking in box~2 must not move the particle out of box~1. Thus \RefEqn{E:moral} applies to $\MPj$; applying \RefEqns{E:sandwich} and \eqref{E:moral} to \RefEqn{E:threebox}, we obtain the latter in its quantum-mechanical form
\begin{equation}\label{E:threeboxQ}
  \Prob[\sz]{\ps{1}{j}}{\MPjOne\wedge\qq}=\frac%
    {\abs{\braket{q}{\p{j}}}^2\abs{\braket{\p{j}}{s}}^2}%
    {\abs{\braket{q}{\p{j}}}^2\abs{\braket{\p{j}}{s}}^2+%
    \big|\Sum{t\neq j}\braket{q}{\p{t}}\braket{\p{t}}{s}\big|^2}.
\end{equation}
(This is a specialization of ABL's (2.4) and (2.5), which, on p.~1413, they extended to incomplete measurements. AV treat it as a new result, their (5).)

If \ket{s} and \ket{q} are such that
\begin{equation}\label{E:threeboxQcondition}
  \braket{q}{\p{1}}\braket{\p{1}}{s}=\braket{q}{\p{2}}\braket{\p{2}}{s}%
   =-\braket{q}{\p{3}}\braket{\p{3}}{s},
\end{equation}
(such as \RefEqn{E:threeboxQconditionEG}), then the retrodictive observation of \p{1} is certain, and the retrodictive observation of \p{2} is certain---the Three-Box ``paradox.''

\subsection{The quantal retrodiction (ABL) formula}
Using $\Prob[s]{y}{x}=\Pr[x]{y}=|\braket{y}{x}|^2$ to express \RefEqn{E:RetroA} in quantum-mechanical terms, we obtain the retrodiction formula for a single complete intermediate observation
\begin{equation}\label{E:ABL}
 \Prob[\sz]{\ps{1}{j}}{\MPs{1}\wedge\qs{2}{k}}=
  \frac%
    {\abs{\braket{\q{k}}{\p{j}}}^2\abs{\braket{\p{j}}{s}}^2}%
    {\Sum{t}\abs{\braket{\q{k}}{\p{t}}}^2\abs{\braket{\p{t}}{s}}^2}.
\end{equation}
(This is the result ABL stated in their (2.4) and (2.5), specialized to a single intermediate complete observation.)

\section{A classical system with pseudo-quantal properties}\label{S:theSystem}%
I present here a strictly non-quantal system which has many statistical properties similar to quantum mechanics, and which, in particular, allows the construction of a Three-Box ``paradox.'' (I discuss in greater depth the use of such classical probability systems in \textcite{Kirkpatrick03a}.) Constructed using ordinary playing cards, this system is as distinct from quantum mechanics as is possible. Each card carries two marks, the ``face'' and the ``suit'' (traditionally with names such as King, Queen, Jack, 10, \dots, and Spades, Hearts, Diamonds, Clubs, respectively); these marks will be treated as system variables, \Face (with values \K, \Q, \J) and \Suit (with values \S, \D, \H). I will refer to these variables generically as $P$ and $Q$: $P,\,Q\in\set{\Face,\,\Suit},\;P\neq Q$, with values \set{\p{j}} and \set{\q{k}}, respectively.

\subsection{The classical system}
The system is a deck of playing cards, each card marked with a \Face value and a \Suit value, and a memory containing the name (\emph{not} the value) of the variable of the preceding observation; the content of the memory is denoted \Mem. The deck is divided into two parts, which we call \These and \Others.
\begin{subequations}\label{E:Example}
\begin{equation}\label{E:ExampleA}\text{\parbox{0.8\textwidth}{
 \begin{description}
 \item To prepare the system in the state $P=\p{j}$:
  \begin{enumerate}\renewcommand{\theenumii}{\arabic{enumii}}
  \item Place all cards with $P=\p{j}$ in \These and the remainder of the deck
          in \Others.
  \item Set the memory to the variable name: $\Mem\leftarrow P$.
  \end{enumerate}
 \end{description}
}}\end{equation}
\begin{equation}\label{E:ExampleB}\text{\parbox{0.8\textwidth}{
 \begin{description}
 \item To observe the variable $P$:
  \begin{description}
  \item If $\Mem=P$ (\ie the observation of $P$ is being repeated)
   \begin{enumerate}\renewcommand{\theenumii}{\arabic{enumii}}
   \item Select a card at random from \These.
   \item Report $\p{j}$, the value of the variable $P$ of this card.
   \end{enumerate}
  \item else ($\Mem\neq P$---the preceding observation was not of $P$)
   \begin{enumerate}\renewcommand{\theenumii}{\arabic{enumii}}
   \item Select a card at random from \Others.
   \item Report $\p{j}$, the value of the variable $P$ of this card.
   \item Prepare the system in the state $P=\p{j}$ (as \RefEqn{E:ExampleA}).
   \end{enumerate}
  \end{description}
 \end{description}
}}\end{equation}
\end{subequations}

The variables each take on the same number of values, $V$. Duplicate cards are allowed under the restriction that each value of each variable appears $N$ times in the deck (so their \emph{a priori} probabilities are equal). For example, we might use the deck \set{(2)\K\S,\K\H,\Q\S,\Q\H} (the ``(2)'' indicates \emph{two} $\K\S$ cards): $V=2$ and $N=3$.

\subsection{The incomplete observation}
In order to create a classical Three-Box-type system, we must make a partial, or incomplete, observation. Fortunately, the partial observation $\Mp{P}{j}=\p{j}\vee \p{j}^{\sim}$ is easily implemented in our example system: To prepare the system in the state $\p{j}^{\sim}$, we simply follow the instructions literally, placing every card with $P$-value $\p{j}^{\sim}$ (\ie every card which satisfies $P\neq \p{j}$) in \These, and all the others (all of which satisfy $P=\p{j}$) in \Others. The observation of $P$ under the manifestation rule \Mp{P}{j} is accomplished exactly as before: we report ``$\p{j}$'' if the card's value of $P$ is $\p{j}$, and ``$\p{j}^{\sim}$'' if the card's value of $P$ is \emph{not} $\p{j}$; for the purpose of the test ``$\Mem=P$,'' a variable is considered the same variable whether partially or fully observed.

Note from \RefEqn{E:Stability} that \Mp{P}{j} is stable (as we required of the quantal manifestation). Formulas for this system's probabilities are given in Appendix \ref{S:Formulas}.

\subsection{The classical Three-Box ``paradoxical'' system}
We can now create a classical ``Three-Box paradox'' example: Take the deck to be \set{(2)\K\H,\,\Q\S,\,\Q\D,\,\J\D,\,\J\S} ($V=3$, $N=2$). We will prepare the system in the state $\Face=\Q$, and filter to (accept only those processes which end in) the final state $\Face=\K$. Let $\Suit=\S$ correspond to box~1, and $\Suit=\D$ correspond to box~2; thus ``opening box~1 (only)'' corresponds to the manifestation $M_{\,\S}^{\sim}=\S\vee{\S}^{\sim}$ (``is the card a \S or not''), and ``opening box~1 (only)'' corresponds to the manifestation $M_{\,\D}^{\sim}=\D\vee{\D}^{\sim}$ (``is the card a \D or not''). This leads to an exact analog to the Three-Box example, but in terms of a deck of cards:

Let us express the deck in the notation $[\These\,|\,\Others]$; then preparation of the state \Q leaves the deck as $\big[\Q\S,\,\Q\D\,\big|\,(2)\K\H,\,\J\D,\,\J\S\big]$. The partial manifestation $M_{\,\S}^{\sim}$ then leads, with probability 1/4, to the occurrence of \S, $\big[\Q\S,\,\J\S\,\big|\,(2)\K\H,\,\Q\D,\,\J\D\big]$, and, with probability 3/4, to the occurrence of ${\S}^{\sim}$, $\big[(2)\K\H,\,\Q\D,\,\J\D\,\big|\,\Q\S,\,\J\S\big]$. Clearly $\Prob[\Q]{\K}{{\S}^{\sim}}=0$, so, by \RefEqn{E:threebox}, $\Prob[\Q^{[0]}]{\S^{[1]}}{{M_{\,\S}^{\sim}}^{[1]}\wedge\K^{[2]}}=1$, and the card is certain to be a \S. A parallel analysis of the partial manifestation $M_{\,\D}^{\sim}$ leads to $\Prob[\Q^{[0]}]{\D^{[1]}}{{M_{\,\D}^{\sim}}^{[1]}\wedge\K^{[2]}}=1$, so the card is certain to be a \D. Thus, in this postselected process, at the intermediate point the card is \emph{both} \S and \D, each with certainty---the ``Three-Box paradox.''

\subsection{Classical interference shown by $\p{j}^{\sim}$  }
The ``paradox'' arises because both \Prob[\Q]{\K}{{\S}^{\sim}} and \Prob[\Q]{\K}{{\D}^{\sim}} vanish. The vanishing of these terms is not trivial: for example, ${\S}^{\sim}$ would seem to include \D, but $\Prob[\Q]{\K}{\D}\neq0$.

The manifestation $M_{\Suit}=\S\vee\H\vee\D$ leads to \S (as before), with a probability of 1/4, to \H, $\big[(2)\K\H\,\big|\,\Q\S,\,\Q\D,\,\J\S,\,\J\D\big]$, with probability 1/2, and to \D, $\big[\Q\D,\,\J\D\,\big|\,(2)\K\H,\,\Q\S,\,\J\S\big]$, with probability 1/4. In this case ``not \S'' is ``$\H\vee\D$,'' the mixture \set{(\H,\,2/3),\,(\D,\,1/3)}; this is easily seen to be $\big[(4)\K\H,\,\Q\D,\,\J\D\,\big|\,(2)\K\H,\,(3)\Q\S,\,(2)\Q\D,\,(3)\J\S,\,(2)\J\D\big]$ (with probability 3/4). Clearly, the mixture corresponding to $\H\vee\D$ differs from the pure state ${\S}^{\sim}$.

It seems to be generally assumed that in classical probability there could be no difference between $\p{1}^{\sim}$ and $\p{2}\vee\p{3}\vee\cdots$; on the other hand, such a difference is known to occur in quantum mechanics, where it is called ``interference,'' and is generally discussed as a mystery specific to quantum mechanics. However, the example presented here demonstrates interference in a classical system: The rules \RefEqn{E:Example} show us that, for every state $s$, $\Pr[s]{{\S}^{\sim}}=\Pr[s]{\H\Or\D}$, suggesting that ${\S}^{\sim}=\H\Or\D$; however (for the deck of this example) $\Prob[\Q]{\K}{{\S}^{\sim}}=0\neq\Prob[\Q]{\K}{\H\Or\D}$, so ${\S}^{\sim}\neq\H\Or\D$. Because quantum interference exhibits exactly this probability behavior, by analogy we say that \H and \D interfere.

If two or three of the ``boxes'' are examined (\ie the complete observation $\S\vee\H\vee\D$ is made), so the results are governed by the probabilities \Prob[\Q^{[0]}]{\p{j}}{{\MP}^{[1]}\wedge\K^{[2]}}, then no \p{j} has a retrodicted probability of 1. The ``Three-Box paradox'' is an interference effect, and that interference is destroyed by the facts-of-the-matter established by examining any two of the boxes---in the original quantal Three-Box setting, in the three-slit interference form (whether atoms or classical waves), \emph{and} in the classical card game.

\section{Not-so-curious statistics}\label{S:Curious}%
The Three-Box example grew out of the ``curious'' example of Albert, Aharonov, and D'Amato (1985) (henceforth ``AAD''); in that work, a pre- and post-selected system is introduced with a contextual behavior which, the authors claim, contradicts the theorems of Gleason and Kochen and Specker.

This conclusion is unwarranted; it arises out of ambiguity regarding the nature, complete or partial, of the intermediate manifestation.

AAD's system has an observable $X$ with eigenstates \set{\ket{x_j}}, an observable $A$ with an eigenstate $\ket{a}=(\ket{x_1}+\ket{x_2})/\sqrt{2}$, and an observable $B$ with an eigenstate $\ket{b}=(\ket{x_2}+\ket{x_3})/\sqrt{2}$. The process is preselected at the time $t_0$ for $A=a$ and postselected at $t_2$ for $B=b$. According to the ABL retrodiction formula, an observation of $X$ at $t_1\in (t_0,\,t_2)$ yields $X=x_2$ with certainty; thus $\Pr{X=x_1}=\Pr{X=x_3}=0$ at $t_1$. (Hence, the authors claim, $A=a$, $B=b$, and $X=x_2$ are ``simultaneously well-defined'' throughout the interval $(t_0,\,t_2)$, though they are all pairwise incompatible.) A fourth variable, $Q$, is introduced, with eigenstates
\begin{equation}\label{E:Qvariable}
  \ket{\q{1}}=\alpha\ket{x_1}+\beta\ket{x_3},\quad%
  \ket{\q{2}}=\ket{x_2},\quad%
  \ket{\q{3}}=\beta^*\ket{x_1}-\alpha^*\ket{x_3}.
\end{equation}
Now, according to AAD,
\begin{quote}
something very curious arises. Suppose that $Q$ is observed within the interval $(t_0,\,t_2)$. It might be expected, since $X=x_2$ within that interval, and since $\ket{\q{2}}=\ket{x_2}$, that such an observation will find, with certainty, that $Q=\q{2}$. But that is not so\dots
\end{quote}
Well, whether that is so or not depends entirely on how $Q$ and $X$ are to be observed. These are different variables, but they share an eigenstate (\ket{x_2}=\ket{\q{2}}); because $x_2=q_2$, also $x_2^{\sim}=q_2^{\sim}$. Thus one would expect that observing $\Mp{Q}{2}\equiv\q{2}\vee\q{2}^{\sim}$ would yield the same result for \q{2} as observing $\Mp{X}{2}\equiv x_{2}\vee x_{2}^{\sim}$ would yield for $x_{2}$. And this is exactly right: applying \RefEqn{E:threeboxQ} to this situation,
\begin{equation}\label{E:AADPartial}
 \Prob[a^{[0]}]{x_2^{[1]}}{{\Mp{X}{2}}^{[1]}\wedge b^{[2]}}=1%
 =\Prob[a^{[0]}]{q_2^{[1]}}{{\Mp{Q}{2}}^{[1]}\wedge b^{[2]}}.
\end{equation}
On the other hand, $\q{1}\vee\q{3}$ is physically different from $x_1\vee x_3$, so the result for \q{2} when observing $M_Q\equiv\q{1}\vee\q{2}\vee\q{3}$ may well differ from the result for $x_{2}$ when observing $M_X\equiv x_1\vee x_2\vee x_3$. Again, this is the case; applying \RefEqn{E:ABL}, we find
\begin{equation}\label{E:AADComplete}
 \Prob[a^{[0]}]{x_2^{[1]}}{M_X^{[1]}\wedge b^{[2]}}=1,\quad\text{but}\quad%
 \Prob[a^{[0]}]{q_2^{[1]}}{M_Q^{[1]}\wedge b^{[2]}}<1.
\end{equation}

AAD inappropriately used the complete observation $M_Q$ rather than the partial observation \Mp{Q}{2}; this led them to \RefEqn{E:AADComplete} rather than to \RefEqn{E:AADPartial}, and hence to an unwarranted sense of ``curiousness'': AAD continue
\begin{quote}
But that is not so: Albeit $\braket{a}{x_3}=\braket{b}{x_1}=0$, yet $\braket{a}{q_1}\neq0$ and $\braket{b}{q_1}\neq0$. Consequently, albeit $\Pr{x_1}=0$ and $\Pr{x_3}=0$ within that interval, $\Pr{q_1}\neq0$ there.
\end{quote}

An ambiguity in the ABL expression for the retrodiction probability led AAD to this confusion: what we have expressed as \Prob[\sz]{\ps{1}{j}}{M_P^{[1]}\wedge\qs{2}{k}}, ABL denoted $p(\p{j}\,/\,s,\,\q{k})$ (and AAD denoted merely $P(\p{j})$), leaving out any mention of the details of the manifestation at the intermediate observation. But, as we have just seen, it is necessary to take into account the degree of completeness of the manifestation, \emph{even of those values not under direct consideration}.

The obvious lesson: use a complete, unambiguous notation which cannot fail to call attention to this need.

\section{The values of a pair of incompatible variables in the interval between their
observations}% 
ABL suggested that, if a system were prepared in the state $P=\p{j}$ at time $t_1$ and observed at time $t_2>t_1$ in the state $Q=\q{k}$, then \emph{both} $P$ and $Q$ are sharp at all times $t\in(t_1,\,t_2)$, so ``we are led into assigning the state [\ket{\q{k}}] to the period of time \emph{preceding} the observation of [$Q$] yielding the eigenvalue [\q{k}].'' This is based on the fact that, according to \RefEqn{E:RetroA}, an observation of $P$ at any such time $t$ would yield, with certainty, the value \p{j}, but also an observation of $Q$ at any such time $t$ would yield, with certainty, the value \q{k}. (Of course, \RefEqn{E:RetroA} allows for only \emph{one} such observation within that time interval.) A considerable controversy has arisen (\cf \cite{Vaidman99b} and \cite{Kastner99b}, and citations within each) regarding exactly what such ``counterfactual'' sharpness might mean.

It is interesting to examine this claim within our card system. Prepare the system in the state \K at time $t_1$, and at time $t_2>t_1$ observe \Suit and select the occurrence if and only if $\Suit=\H$. Then, exactly in the sense of ABL, \emph{both} \K and \H are sharp throughout the interval $(t_1,\,t_2)$.  But we can see inside this system: if during $(t_1,\,t_2)$ either no observation, or an observation of \Face, is made, then $\Face=\K$ throughout but \Suit has no value until $t_2$; on the other hand, if at time $t\in(t_1,\,t_2)$ an observation of \Suit is made, then during the interval $(t_1,\,t)$ \K is sharp and \Suit has no value, and during the interval $(t,\,t_2)$ \H is sharp and \Face has no value. That the observation of \Suit has the definite result of \H implies nothing about its prior-to-observation value; the final selection simply throws away all occurrences of \D and \S. There is no warrant for the claim that these variables have simultaneously sharp values.

Note also how inappropriate it is to apply the term ``measurement'' to these procedures: clearly, no value of $Q$ exists at time $t_1$ to be measured; even the term ``observation'' is misleading. In fact, an interaction of a specific nature has occurred which has brought to the variable $Q$ a value; it is this that I have called ``manifestation.'' It is interesting to read the Vaidman--Kastner discussion (as well as the the much older and more extensive literature concerning quantum ``reality'') with this example in mind.

\section{Conclusion}
The Three-Box example arose in a quantum setting, and was taken (somewhat uncritically) to be another example of the ``bizarre'' nature of quantum mechanics. The restatement of the example as a three-slit atomic Young system shows it to be a straightforward example of quantal interference. But exemplified in a perfectly ordinary setting, a deck of cards without a quantum in sight, the Three-Box phenomenon becomes merely an interesting phenomenon of ordinary probability systems, exhibited by a quantum-mechanical system \emph{qua} probability system, in no way a quantal phenomenon---hence my characterization of Vaidman's comment as ``much ado about nothing.''

Quantum mechanics provides only the probability of events; it is generally agreed that there is no underlying mechanism (and certainly, if there is, we know nothing of it). Thus all we know of the quantum Three Box example is this: A system is prepared in a certain way; one or another of two partial observations of a system variable is made; a final observation is used to filter out a single outcome; under a certain choice of the preparation and the final filter, each of the intermediate observations has a sharp outcome.  Quantum mechanics tells us nothing more. This card game satisfies the same probabilistic description; every probabilistic quality is common to both systems, and the quantitative differences in probabilities fail to provide a qualitative distinguishing feature.

Their significant difference between them is that the card system has a known underlying mechanism which can be analyzed and understood. First, such analysis shows no \emph{ad hoc} devices---the card system follows its own internal logic consistently to the Three-Box-like result. Second, analysis of the interior state of this classical system makes it clear that there is no ontic significance to these retrodictively sharp-if-and-when-observed values---the ``observation'' actually \emph{brings the value into existence}. Thus, the mere fact that a value is statistically definite does not imply that the observable ``has'' that value (in any reasonable sense of ``to have''): \emph{A sharp value is not necessarily a possessed value}. This ``failure of realism'' in an obviously real system undercuts metaphysical concerns regarding such failure in quantum mechanics and supports the conclusion that the Three-Box example has no ontic significance regarding possessed values.

%%% APPENDIX %%%%%%%%%%%%%%%%%%%%%%%%%%%%%%%%%%%%%%%%%%%%%%%%%%%%%%%%%%%%%%%%%%%%%%%%%%
\appendix
\setcounter{equation}{0}

\section{Notation}\label{S:Notation}%
Throughout this paper, $P$ and $Q$ represent distinct system variables with possible values \set{\p{j}} and \set{\q{k}}, respectively. The proposition that a variable has a certain value, \eg ``$P=\p{j}$,'' is abbreviated with the variable's value, ``$\p{j}$.'' A general preparation of the system will be denoted $s$; we write probability expressions with the preparation as a subscript: \Pr[s]{\p{j}} is the probability of the proposition $P=\p{j}$ after the preparation in the state $s$.

The conditional probability (probability conditioned on an occurrent fact), defined by%
\footnote{%
Disjunction (``or'') is indicated by $\vee$; conjunction (``and''), by $\wedge$; negation (``not'') by $\sim$.
} %
\begin{equation}\label{E:Cond}
 \Prob[s]{b}{a}=
 \begin{cases}
   \Pr[s]{a\wedge b}/\Pr[]{a}&\Pr[]{a}>0\\
   \text{undefined}&\text{otherwise},
 \end{cases}
\end{equation}
is the probability of the truth of the proposition $b$ given that the fact stated by the proposition $a$ occurs.

The set of propositions \set{a_j} is \emph{disjoint} iff, whenever all \set{a_j} take on values, $a_j\wedge a_{j'}=F,\;j\neq j'$. A disjoint set satisfies $\Sum{t}\Pr[s]{a_t}=\Pr[s]{\Orj{t}{a_t}}$ for all preparations $s$.

The set \set{a_j} is \emph{complete} iff, whenever all \set{a_j} take on values, $\Orj{t}{a_t}=\text{T}$; hence, a disjoint, complete set satisfies $\Sum{t}\Pr[s]{a_t}=1$ for all preparations $s$.

An \emph{event} is an occurrence at which at least one variable of the system takes on a value randomly; this is brought about by a physical interaction of the system with its exterior. \emph{Which} variable takes on a value randomly depends on the details of the physical interaction, or \emph{manifestation}; at each event, then, a particular variable is manifested. Manifestation will be indicated as a condition of the probability (as discussed in \RefSec{S:ClassDeriv}).

The ordinal position of an event in a sequence of events will be denoted by a superscript in brackets: The event $E$ followed by the event $F$ is denoted $E^{[1]}\wedge F^{[2]}$. However, when the terms in the probability expressions are in the ``natural'' order and no ambiguity arises, the sequence ordinals will be dropped; thus \Pr[s]{E} always means \Pr[\sz]{E^{[1]}}, and \Prob[s]{F}{E} always means \Prob[\sz]{F^{[2]}}{E^{[1]}}.

\section{Stability}%
In quantum mechanics a proposition $p$ is represented by the projector \Proj{p}. If the manifestation \Mp{P}{k} is stable%
\footnote{%
Stability is implied by, but weaker than, Wigner's ``morality'' (\cf \textcite{GoldbergerWatson64}).} % 
in the sense that
\begin{equation}\label{E:Stability}
 \Prob[\ps{0}{j}]{{\p{k}^{\sim}}^{[1]}\wedge\ps{2}{j'}}{{\Mp{P}{k}}^{[1]}}=%
    \KDelta{j}{j'}(1-\KDelta{j}{k})
\end{equation}
then, applying the Wigner ``sandwich'' formula for the probability of successive non-disjunctive events,
\begin{equation}\label{E:sandwich}
  \Pr[s]{\ps{1}{}\wedge\qs{2}{}}%\equiv\Prob[s]{\q{}}{\p{}}\Pr[s]{\p{}}%
       =\Trace{\bRho[s]\Proj{\p{}}\Proj{\q{}}\Proj{\p{}}},
\end{equation}
we obtain
\begin{equation}
 \abs{\bra{\p{j}}\Proj{\p{k}^{\sim}}\ket{\p{j'}}}^2=\KDelta{j}{j'}(1-\KDelta{j}{k}).
\end{equation}

In order that \Proj{\p{k}^{\sim}} be a projector, the diagonal elements of its matrix must be the positive square roots, hence
\begin{equation}\label{E:moral}
 \Proj{\p{k}^{\sim}}=\Sum{t\neq k}\Proj{\p{t}}=\One-\Proj{\p{k}}.
\end{equation}

\section{Probabilities of the card system}\label{S:Formulas}%
The example of \RefSec{S:theSystem} satisfies the following probability expressions:
\begin{subequations}\begin{align}
 &\Prob[s]{\cdot}{\p{j}}=\Pr[\p{j}]{\cdot},&&\text{ if }\;\Pr[s]{\p{j}}\neq0&\\
 &\Prob[\q{j}]{\cdot}{\p{k}^{\sim}}=\Pr[\p{k}^{\sim}]{\cdot},\;%
       &&\text{if}\;\Pr[\q{j}]{\p{k}^{\sim}}\neq0&\\
 &\Prob[\p{j}]{\cdot}{\p{k}^{\sim}}=\Pr[\p{j}]{\cdot},&&\text{ if}\;j\neq k&
\end{align}\end{subequations}
\begin{subequations}\begin{align}
 \Pr[\p{j}]{\p{k}}&=\KDelta{j}{k}&
 \Pr[\p{j}]{\q{k}}&=\dfrac{N-N(\p{j}\cdot\q{k})}{N(V-1)}\\
 \Pr[\p{k}^{\sim}]{\p{j}}&=(1-\KDelta{j}{k})\dfrac{1}{V-1}&
 \Pr[\p{j}^{\sim}]{\q{k}}&=\dfrac{N(\p{j}\cdot\q{k})}{N}
\end{align}\end{subequations}
\begin{equation}
   \Pr[s]{\p{j}^{\sim}}=1-\Pr[s]{\p{j}}
\end{equation}

%%%%%%%%%%%%%%%%%%%%%%%%%%%%%%%%%%%%%%%%%%%%%%%%%%%%%%%%%%%%%%%%%%%%%%%%%%%%%%%%%%%%%%%%
%% BIBLIOGRAPHY %%%%%%%%%%%%%%%%%%%%%%%%%%%%%%%%%%%%%%%%%%%%%%%%%%%%%%%%%%%%%%%%%%%%%%%%
%%%%%%%%%%%%%%%%%%%%%%%%%%%%%%%%%%%%%%%%%%%%%%%%%%%%%%%%%%%%%%%%%%%%%%%%%%%%%%%%%%%%%%%%
 \renewcommand{\refname}{\sc References}%
 \footnotesize%
%%%%%%%%%%%%%%%%%%%%%%%%%%%%%%%%%%%%%%%%%%%%%%%%%%%%%%%%%%%%%%%%%%%%%%%%%%%%%%%%%%%%%%%%
% \bibliographystyle{myapalike}
% \bibliography{JAbbrevs,QM}

%%%%%%%%%%%%%%%%%%%%%%%%%%%%%%%%%%%%%%%%%%%%%%%%%%%%%%%%%%%%%%%%%%%%%%%%%%%%%%%%%%%%%%%%
%%% END DOCUMENT %%%%%%%%%%%%%%%%%%%%%%%%%%%%%%%%%%%%%%%%%%%%%%%%%%%%%%%%%%%%%%%%%%%%%%%
%%%%%%%%%%%%%%%%%%%%%%%%%%%%%%%%%%%%%%%%%%%%%%%%%%%%%%%%%%%%%%%%%%%%%%%%%%%%%%%%%%%%%%%%
\end{document}